\documentstyle[editedvolume,namedreferences]{crckapb}

\begin{opening}
\title{ASTROPHYSICS OF THE SOFT GAMMA REPEATERS
AND ANOMALOUS X-RAY PULSARS}
\author{Christopher Thompson}
\institute{University of North Carolina,\\
Department of Physics and Astronomy,\\
Chapel Hill, NC 27599}

\end{opening}

\runningtitle{SOFT GAMMA REPEATERS AND ANOMALOUS X-RAY PULSARS}

\begin{document}

I summarize the recent advances in our understanding of
the Soft Gamma Repeaters: in particular their spin behavior,
persistent emission and hyper-Eddington outbursts. 
The giant flares on 5 March 1979 and 27 August 1998
provide compelling physical evidence for magnetic fields stronger than
$10\,B_{QED} = 4.4\times 10^{14}$ G, consistent with the
rapid spindown detected in two of these sources.  The persistent
X-ray emission and variable spindown of the 6-12 s Anomalous X-ray Pulsars
are compared and contrasted with those of the SGRs, and the case made
for a close connection between the two types of sources.  
Their collective properties point to the existence of {\it magnetars}:
neutron stars in which a decaying magnetic field
(rather than accretion or rotation) is the dominant source of energy for
radiative and particle emissions.  Observational tests of the magnetar
model are outlined, along with current ideas about the trigger of SGR 
outbursts, new evidence for the trapped fireball model, and the influence of
QED processes on X-ray spectra and lightcurves.  A critical 
examination is made of coherent radio emission from bursting strong-field
neutron stars.  I conclude with an overview of the
genetic connection between neutron star magnetism and the violent
fluid motions in a collapsing supernova core.

\section{Introduction}

During the last 30 years, a comfortable picture of the Galactic neutron
star
population emerged:  neutron stars are born with largely dipolar
magnetic fields
of $\sim 10^{11}-10^{13}$ G,  which do not decay significantly unless
the star accretes upwards of $\sim 0.1\,M_\odot$ from a binary
companion.
This picture is based on observations of neutron stars whose pulsed
emissions are powered either by rotation, or by accretion.  In the first
case, there are strong selection effects against observing radio
pulsations
from a star whose dipole magnetic field is much stronger than 
the quantum electrodynamic value $B_{QED} = m_e^2c^3/e\hbar = 
4.4\times 10^{13}$ G.  At a fixed age, the spin period $P \propto
B_{dipole}$ -- after the magnetic dipole torque has pushed $P$ well
above its initial
value -- and the spindown luminosity $I\Omega\dot\Omega \propto
B_{dipole}^{-2}$.
The radio pulsations are also expected to be beamed into an increasingly
narrow
solid angle, a dramatic example being the `new' 8.5 s PSR J2144-3933
(Young, Manchester, \& Johnston 1999). 
The upper envelope of the distribution of measured pulsar
dipole fields has, nonetheless, increased significantly with the recent
discovery of PSRs J1119-6127 and J1814-1744, the second of which is
inferred to have a polar field in excess of $10^{14}$ G (Camilo et al.
2000).
The apparent paucity of neutron stars with $B_{dipole} > B_{QED}$ in
accreting systems places tighter constraints on their birth rate {\it
if} they have the
same distribution of natal kicks as ordinary radio pulsars.

Although strong by terrestrial standards, a $\sim 10^{12}$ G magnetic
field is, in a dynamical sense, quite weak.  It
contributes only $\sim 10^{-9}$ of the hydrostatic pressure
(when the effects of proton superconductivity in the
stellar core are taken into account).  Much stronger magnetic fields
can be generated by vigorous convective motions in a supernova core,
$B \sim 10^{15}$ G (Thompson \& Duncan 1993, hereafter TD93). 
Substantial evidence has accumulated in recent
years\footnote{Due to the very unfortunate absence of Jan van Paradijs
and Chryssa Kouveliotou,
this review covers the phenomenology of the Soft Gamma Repeaters and
Anomalous X-ray Pulsars, as well as theoretical aspects of
strong-field neutron stars.  It combines two recent summaries
of the bursting behavior of the SGRs (Thompson 2000b) and the 
persistent emission and spindown of the SGRs and AXPs (Thompson 2000c).
See Norris et al. (1991) for a review of
the early literature on the SGRs, and Frail (1998) and
Mereghetti (2000) for a more complete review of radio-silent neutron
stars and the AXPs.} for
neutron stars whose much stronger magnetic fields ($B_{dipole}
\sim 10\,B_{QED} = 4.4\times 10^{14}$ G) decay significantly on a very
short timescale ($\sim 10^4$ yr).  These {\it magnetars} were predicted
to spin down much more rapidly than ordinary radio pulsars, and
should be elusive (although not necessarily impossible to detect) as
pulsed radio sources.  They have been most cogently
associated (Duncan \& Thompson 1992, hereafter DT92; Paczy\'nski 1992;
Thompson \&
Duncan 1995, hereafter TD95) with the Soft Gamma Repeaters:
a small peculiar class of neutron stars that emit extremely luminous and
hard X-ray and gamma-ray bursts.  The growing group of Anomalous X-ray Pulsars
(Mereghetti 2000) may be closely related (Thompson \& Duncan 1996,
hereafter TD96).

 The defining property of a magnetar is that
its decaying magnetic field outstrips its rotation as a source of energy
for
X-ray and particle emission ---  by some  two orders of magnitude
if the core field is as strong as $\sim 10^2\,B_{QED}$ (TD96).
The observational signatures of this decay
include persistent X-ray and particle
emissions and, if $B > (4\pi\theta_{max}\mu)^{1/2} =
2\times 10^{14}\,(\theta_{max}/10^{-3})^{1/2}$ G, sudden outbursts
triggered
by fractures of the rigid crust.  (Here $\mu$ is the shear modulus
and $\theta_{max}$ the yield strain in the deep crust.)

\section{Soft Gamma Repeaters}

The SGRs are best known for two giant flares
on March 5, 1979 and August 27, 1998 (from SGR 0526-66 and SGR 1900+14
respectively).  These remarkable 
bursts, separated by almost 20 years, are nearly carbon
copies of each other.  They released $\sim 4\times 10^{44}$ and
$\sim 1\times 10^{44}$ erg in X-rays respectively (Cline 1982;
Hurley et al. 1999a;
Feroci et al. 1999; Mazets et al. 1999b and references therein)
and had very similar and striking morphologies.  Each was initiated by a
short and intense ($t \sim 0.2-0.5$ s) spike.  The luminosity of
this spike exceeded the classical Eddington luminosity -- above which
the outward force due to electron scattering exceeds the attractive
force of gravity:
$L_{edd} \simeq 2\times 10^{38}$ erg s$^{-1}$ for a $1.4\,M_\odot$
neutron
star --  by a factor $3\times 10^6-10^7$ in the case of the March 5
event
(Fenimore, Klebesadel, \& Laros 1996).  The ensuing softer emission,
which
lasted $200-400$ s and radiated somewhat more energy, had a much more
stable temperature even though its luminosity exceeded $\sim
10^4\,L_{edd}$.
In the March 5 flare, this tail had a striking 8-second periodicity of a
very large amplitude, which was inferred to be the rotation period of
SGR 0526-66.  The August 27 event exhibited a similar
5.16 s periodicity of an even larger amplitude. 

These SGRs are more generally characterized by short ($\sim 0.1$ s)
X-ray outbursts with luminosities as high as $10^4\,L_{edd}$.
The statistics of the these short bursts have some
intriguing similarities with earthquakes and Solar flares (Cheng et al. 1996).
According to the most recent analyses,
the bursts of SGR 1806-20 ({G\"o\u{g}\"u\c{s}} et al. 1999b) and 
SGR 1900+14 ({G\"o\u{g}\"u\c{s}} et al. 1999a) have a power-law
distribution of energies $dN/dE \propto E^{-1.6}$ that extends over 
some 4-5 decades.  The 
distribution of waiting times is lognormal, peaking at $\sim 1$ day
(Hurley et al. 1996; {G\"o\u{g}\"u\c{s}} et al. 1999a,b).
In the case of SGR 1806-20, continuous monitoring in 1983 showed that the
burst fluence accumulates in a piecewise-linear manner while the source
is active (Palmer 1999).  This indicates the existence of multiple
active
regions, internal to the star.

The four known SGRs are also persistent X-ray sources of luminosity
$10^{35}-10^{36}$ erg s$^{-1}$ (Rothschild, Kulkarni,
\& Lingenfelter 1994; Murakami et al. 1994; Hurley et al. 1999b;
Woods et al. 1999b).  Although the time average of the bursting
emission is uncertain due to the intermittency of SGR activity, it
is in order of magnitude comparable to the persistent output.
Periodicites  have been convincingly detected in two sources:  
$P = 7.47$ s for SGR 1806-20 (Kouveliotou et al.
1998);  and $P = 5.16$ s for SGR 1900+14 (Hurley et al. 1999c).
This measurement of the spin preceded the August 27 event
in the case of SGR 1900+14, and agreed with the periodicity
detected in the giant outburst.  This narrow clustering of
persistent luminosities and especially of spin periods 
in the SGR sources provides an important clue to the energy source
that powers them.

Evidence that SGR 0526-66 is a young neutron star comes from
its positional coincidence with a supernova remnant
N49 (age $\sim 7\times 10^3$ yr) in the LMC.  If the star has 
been spun down by a vacuum magnetic dipole torque to a period of 8 s,
then one infers a polar dipole field of $6\times 10^{14}\,
(t/10^4~{\rm yr})^{-1/2}$ G (Duncan \& Thompson 1992).
Additional evidence that the SGRs are young neutron stars comes from
the association of the other three -- with varying degrees of certainty 
-- with supernova remnants (Kulkarni \& Frail 1993; Vasisht et al. 1994;
Hurley et al. 1999b; Woods et al. 1999b) or regions of massive star
formation (Fuchs et al. 1999).

An important observational breakthrough, which supports this
interpretation, came with the recent detection
of rapid spindown in SGRs 1806-20 and 1900+14.  Both sources have
(coincidentally) nearly the same characteristic age of $P/\dot P =
3000$ yr (Kouveliotou et al. 1998, 1999;  Marsden, Rothschild, \& Lingenfelter
1999; Woods et al. 1999c; Woods et al. 2000).  The inferred polar
magnetic field strength exceeds $10^{15}$ G in both cases, in the
simplest model of an orthogonal vacuum rotator.  
Deviations from this torque behavior are outlined in \S 5.2.
%Nonetheless, the measured spindown luminosity $I\Omega\dot\Omega$
%is smaller by two orders of magnitude than the persistent X-ray
%luminosity.
%In this situation, the inferred dipole field of both SGRs would be
%reduced by a factor of $\sim 4$ to $4\times 10^{14}$ G
%if the star were a persistent source of Alfv\'en waves and particles
%with a luminosity comparable to $L_X \sim 10^{35}$ erg s$^{-1}$
%(Thompson \& Blaes 1998; Harding, Contopoulos, \& Kazanas 1999;
%Thompson et al. 1999).  

\section{Anomalous X-ray Pulsars}

The Anomalous X-ray Pulsars constitute a separate group of a 
half-dozen neutron stars that have been detected through their
persistent X-ray
pulsations but have not (yet) been observed to burst  (Mereghetti \& Stella
1995; Duncan \& Thompson 1995;  Van Paradijs, Taam, \& Van den Heuvel
1995; Mereghetti 2000).  The AXPs share with the SGRs
remarkably similiar persistent X-ray luminosities
($L_X \sim 3\times 10^{34}-10^{36}$ erg s$^{-1}$),
spin periods ($P \sim 6-12$ s),  and characteristic ages
($P/\dot P \sim 10^3 - 10^5$ yr).  They are all consistently
spinning down.  At least three are associated with young supernova remnants.  

This overlap between the SGRs and AXPs in a {\it three}-dimensional
parameter space would be surprising if the two classes of sources
were powered by fundamentally different energy sources.  Accretion
has been suggested for the AXPs -- either in a low mass X-ray binary
(Mereghetti \& Stella 1995) or through a fossil disk (Van Paradijs et al.
1995).  In that case, one infers $B_{dipole} \sim 10^{11}-10^{12}$ G.
However, the need for an active energy source in the SGRs,
almost certainly involving 
a strong magnetic field, combined with the variability
of the X-ray output of some AXPs, led to the suggestion that
the AXPs were also powered by a decaying magnetic field, and were
spinning down essentially by a vacuum magnetic dipole torque (TD96;
Kouveliotou et al. 1998).  In that case, the estimated dipole fields
rise to $\sim 10^{14}-10^{15}$ G.  
The possibility remains that some AXPs are strongly magnetized
but otherwise passively cooling neutron stars (Heyl \& Hernquist 1997a,b).  
The SGRs are highly intermittent as burst sources
(activity in SGR 0526-66 was only detected during the years 1979 to 1983);
this provides a hint that some AXPs have experienced X-ray outbursts
in the past and will burst in the future.  Combining both classes of
sources, one roughly estimates the net birth rate
of SGRs/AXPs as $\sim 1\times 10^{-3}$ per year (Thompson et al. 2000a,
hereafter T00).

The identification of AXPs with inactive magnetars
resolves some of the problems associated
with accretion models.  There is no evidence for binarity in
any of these sources (Kaspi, Chakrabarty, \& Steinberger 1999; Mereghetti
2000).  If their X-ray emission
were powered by accretion from a low-mass binary companion, then the
long orbital
evolution time --  $10^7-10^8$ yr due to gravity wave emission --
combined with the presence of a few AXPs inside SNR of age $\sim 10^4$
yr would imply far
more sources in the Galaxy than are in fact observed (TD96).  The
fossil disk model appears to be on firmer grounds, and is testable
by searches for optical/UV emission reprocessed by the disk from
the central X-ray continuum.  (See Perna, Hernquist, \& Narayan 2000
for detailed modelling.)  The X-ray spectra of the AXPs are peculiarly
soft by the standards of X-ray binaries, which suggests that if they are
accreting then their magnetic fields are weaker than $10^{12}$ G.  
Fields that weak can barely provide the spindown torque measured in the AXPs
1E 1048.1+5937 and 1E 1841-045 (Li 1999).  

Interest in the connection between the AXPs and radio pulsars has been
raised by
the recent discovery of PSR J1814-1744, which is positioned near the AXP
1E 2259+586 in the $P-\dot P$ plane (Camilo et al. 2000).   It should
first be noted that the spindown
age of 1E 2259+586 ($P/2\dot P \sim 2\times 10^5$ yr) is $> 10$ times
the age of the
supernova remnant CTB 109 near whose center it sits.  Since all the
other AXPs and SGRs for
which this comparison can be made have {\it shorter} spindown ages, it
seems likely
that the spindown torque of 1E 2259+586 has decayed, and
that over most of its history this AXP sat a factor of $\sim 10$ higher
in
the $P-\dot P$ plane.  Several effects -- alignment, field decay, or a
previous phase of accelerated spindown -- could explain this observation
in the magnetar model (T00). 

The strong-field pulsar J1814-1744 is not a conspicuous X-ray source,
being at least 10-100 times weaker than the AXPs
(Pivovaroff, Kaspi, \& Camilo 2000).  More generally, there appears
to be a bifurcation in the level of internally-generated dissipation
between the SGRs (and AXPs) and radio pulsars of a similarly young age.
A critical parameter which could explain this bifurcation is provided
by the internal (e.g. toroidal) magnetic field (TD96).  Above a flux
density of $\sim 100\,B_{QED}$ -- corresponding to 5-10 times the dipole
fields inferred from spindown -- there is a rapid increase
in the rate of ambipolar diffusion through the degenerate core
(see also \S 7).  

\section{SGR Outbursts:  Physical Diagnostics}

The strength of the magnetic field is probably the most important
parameter to be determined in the SGR and AXP sources.  In this section, 
we review how the extreme properties of SGR outbursts directly point 
to flux densities higher than $10\,B_{QED} = 4.4\times 10^{14}$ G in 
these sources.

\subsection{Trigger and Energetics of SGR Bursts}

To power one giant flare like the March 5 and August 27 events, an SGR
must be able to store $\sim 10^{45}$ erg of potential energy.  
This should be compared with the
maximum elastic energy $\sim 10^{44}\,(\theta_{max}/10^{-2})^2$ erg that
can be stored by the crust of a neutron star whose yield strain
is $\theta_{max}$.  It should
be emphasized at this point that a magnetic field
stressing the outer Coulomb lattice of the star actually contains
more available energy ($\delta B^2/4\pi$) than is stored by the lattice
itself
($\sim {1\over 2}\theta^2\mu$ where $\mu$ is the shear modulus).
The ratio is $\sim 4\pi\mu/B^2 = 10^2\,(B/10\,B_{QED})^{-2}$.
Nonetheless, even with this amplification it is not possible for a star
composed of {\it strange quark matter} to retain enough potential magnetic
energy to power the March 5 and August 27 giant flares.
The elastic energy of its crust is smaller than that of a neutron
star by at least four orders of magnitude (Alcock et al. 1986).

Why are the giant flares so rare, when short SGR bursts are relatively
common?  Why is there a gap of more than two orders of magnitude in
fluence?  One possible explanation is that the giant flares involve
a large propagating fracture in the neutron star crust, whereas
the short bursts require only a localized yield of the crustal lattice
(TD95).  A large scale motion of the crust is highly constrained
compared with, e.g.,  a metal sheet, because the crust floats stably on
the neutron star core and is very nearly incompressible.  For this
reason, a large fracture requires the collective and simultaneous
motion of many smaller units.  It is interesting to note, in this regard,
that some earthquake models based on cellular automata show a 
bimodal distribution of events, with a secondary peak at
the highest energies (Carlson, Langer, \& Shaw 1994).

How precisely is energy injected into the magnetosphere of the neutron
star?  The very fast ($\sim$ ms) rise times both of some short SGR
bursts
(Kouveliotou et al. 1987) and the giant outbursts (Fenimore et al. 1996;
Hurley et al. 1999a; Mazets et al. 1999b) point to a localized and
direct
injection of energy.
Indeed, much of the energy that is eventually radiated in the burst may
have been injected on a much shorter timescale than the measured
duration
of the X-ray pulse.  Direct evidence for this behavior comes
from the intense initial spikes of the March 5 and August 27 events,
which released a few tens of percent of the net outburst energy over
$\sim
10^{-3}$ of the duration.  

A magnetic field $B > (4\pi\theta_{max}\mu)^{1/2} \sim 2\times
10^{14}\,
(\theta_{max}/10^{-3})^{1/2}$ G
can fracture the crust, but is far too weak to induce anything but a
horizontal motion.  As a result, energy is injected in the
magnetosphere, in two distinct regions.  The motion will,
in general, have a rotational component that creates tangential
discontinuities in the magnetic field.  A disturbance of the
magnetosphere propagates at the speed of light, which is some 300 times
the shear wave velocity $c_\mu$ in the deep crust.  Thus, reconnection
occurs rapidly, and induces transverse Alfv\'en waves of
frequency $\sim c/R_{NS}$ on the connecting closed loops of magnetic
flux.
These waves can dissipate effectively by  cascading to high wavenumber
through non-linear interactions (Thompson \& Blaes 1998).  Because
they are current-carrying, a minimal density of
electrically charged particles is required to support them, 
which in the giant flares implies a significant 
optical depth to scattering across the
magnetosphere.  The current density (and optical depth) rises
as the cascade moves to higher wavenumber, where it is finally
damped by Compton drag off the electrostatically heated
pairs.  When the rate of transfer of wave energy
exceeds a critical level $10^{42}
\,(\ell/10~{\rm km})$ erg s$^{-1}$ within a volume $\ell^3$, no stable
equilibrium between heating and radiative cooling is possible. 
The plasma runs away to a dense, hot fireball which cools diffusively
on a much longer timescale (TD95; Thompson et al. 2000b).

The second
dissipative region lies much farther out in the magnetosphere.  The
crustal motion (on a horizontal scale $\ell$) can be expected to excite
shear waves of frequency $\nu \sim c_\mu/\ell$, which in turn
couple to magnetospheric Alfv\'en modes at a radius $R_\nu \sim c/3\nu
\sim 100\,\ell$.  This outer excitation may dominate if (for example) the
fracture is buried deep in the crust, and has been identified with two
bursts
from SGR 1900+14 whose hard spectra resemble those of
cosmological GRBs (Woods et al. 1999d).  
%In this second regime,
%where $L_{cas} < L_{crit}$,
%it is possible to maintain a steady balance between elecrostatic
%heating and diffusive radiative cooling.

\subsection{Fracturing vs. Plastic Creep}

Can the magnetar model accomodate two classes of sources with
widely different bursting behavior, but similar levels of
internally-generated
dissipation?  If the magnetic field in the deep crust
exceeds $B_\mu \equiv (4\pi\mu)^{1/2} \sim 6\times 10^{15}$ G,
lattice stresses are not able to compensate a departure from
magnetostatic equilibrium, and the crust must deform plastically
(TD96).  Indeed, the internal flux density above which the magnetic
field is transported through the neutron star core on a timescale of
$\sim
10^4$ yr lies close to this value (TD96; Heyl \& Kulkarni 1998).  This
suggests that a magnetar is capable of two dissipative modes:  one
dominated by brittle fracturing and bright X-ray outbursts, and a second
dominated by plastic creep.
These two modes correspond naturally to the SGRs and
to the Anomalous X-ray Pulsars.  In principle, both modes can operate
simultaneously in the same star if its magnetic field is inhomogeneous.

\subsection{Hard Spikes of the March 5 and August 27 Events}

The initial spikes of the two giant outbursts had all the appearance
of an expanding $e^\pm$ fireball carrying $\sim 10^{44}$ erg.
(In the case of the March 5 event, $L \sim 3\times 10^6- 10^7$ $L_{edd}$
and
$T \sim 500$ keV:  Mazets et al. 1999b; Fenimore et al. 1996.)
The peak luminosity is intermediate, on a logarithmic scale, between
that of a thermonuclear X-ray flash and the bright $\gamma$-ray
fireballs that are observed at cosmological distances.
  If the fireball contained comparable energy in radiation and in
the rest energy of (baryonic) matter, then its duration could
be expressed in terms of the radius $R(\tau_{es} = 1) \geq
(E\sigma_T/4\pi m_pc^2)^{1/2}$ of the
scattering photosphere as
$\Delta t \sim R(\tau_{es} = 1)/c \sim 2(E/10^{44}~{\rm erg})^{1/2}$
s, about 10 times the observed value.  We conclude that the initial
fireball must have in fact expanded relativistically, and was powered
by a very clean energy source.

The most obvious candidate is a magnetic field that experiences a sudden
rearrangement.   On energetic grounds, the (external) magnetic field
must exceed $\sim 10B_{QED}$ to power $\sim 10^2$ giant outbursts
over $\sim 10^4$ yr.
One might consider a hybrid model in which the energy is initially
released inside the
neutron star, in the form of crustal shear waves or torsional Alfven
waves in the
liquid core.  As we now show, fields of comparable strength are
required to transmit this energy to the magnetosphere.  The large
output of the
giant outbursts requires a large scale for this energy release, and
hence a low frequency for the excited mode.  For example, a fracture of
dimension $\ell \sim 1$ km (a conservative lower bound) will
excite a shear wave of frequency  $\nu \sim 10^3\,(\ell/1~{\rm
km})^{-1}$ Hz.
The resulting harmonic displacement $\xi$ of the crust will in turn
excite oscillations of the dipolar magnetic field lines at a radius
$R_\nu \sim c/3\nu \sim 10^7\,(\nu/{\rm kHz})^{-1}$ cm.  Because only a
narrow bundle of the outer field is excited, the outward wave luminosity
is a
steep function of $\xi$, $dE_{wave}/dt \simeq {1\over 2}B_{dipole}^2
R_{NS}^2
c\,(2\pi\xi \nu/c)^{8/3}$, or
\begin{equation}
{dE_{wave}\over dt} \simeq  2\times 10^{44}\,
\left({B_{dipole}\over 10\,B_{QED}}\right)^2\,
\left({\xi\over 0.1~{\rm km}}\right)^{8/3}\,
\left({\nu\over 10^3~{\rm Hz}}\right)^{8/3}\;\;\;\;\;\;{\rm erg~s^{-1}},
\end{equation}
(Thompson \& Blaes 1998).  For example, an elastic
distortion of the crust of energy $\sim 10^{44}$ erg corresponds to $\xi
\sim 10^{-2}\, R_{NS} \sim 0.1$ km, and the luminosity approaches
$10^7\,L_{edd}$ only if $B_{dipole} \sim 10^{15}\,(\nu/10^3\,{\rm
Hz})^{-4/3}$
G! 

Nonetheless, the short $\sim 0.2-0.5$ s duration of the initial hard
spikes of the March 5 and August 27 outbursts  provides direct evidence
that internal (rather than external) magnetic stresses trigger these
giant outbursts.  A $10^{15}$ G magnetic field will move
the core material at a speed $\sim B/\sqrt{4\pi\rho}$ through a distance
$10$ km in that period of time.   By contrast, the fireball resulting
from a sudden unwinding of the external field would last only
$\sim R_{NS}/c \sim 10^{-4}$ s (TD95).

\subsection{Oscillatory Tails of the March 5 and August 27 Events}

After the initial hard spike, each of the two giant outbursts
released an even greater amount of energy in an extended oscillatory
tail.
The temperature during this phase was much more stable in spite of
the hyper-Eddington flux, $L/L_{edd} \sim 10^4$ (e.g. Mazets et al.
1999b).
These observations suggest that a significant fraction of the initial
burst
of energy was trapped on closed magnetic field lines,
which implies a strong lower bound to the
surface dipole magnetic field, $B_{dipole} > 2\times 10^{14}\,
(E_{fireball}/10^{44}~{\rm erg})^{1/2}(\Delta R/10~{\rm km})^{-3/2}$\break
$[(1~+~\Delta R/R_{NS})/2]^3$ G (TD95; Hurley et al. 1999a).

The density of this trapped energy is so high that it must form
a thermal fireball, composed of $e^\pm$ pairs and $\gamma$-rays, at a very
high temperature of $\sim 1$ MeV. The optical depth to scattering across 
this plasma bubble is huge, approaching $\sim 10^{10}$.  It is
clear that the plasma cannot cool by simple radiative diffusion from its
center:  that would take $\sim 10^3-10^4$ times the observed burst
duration.
The bubble cools instead as a sharp temperature gradient develops just
inside its outer boundary, and this boundary propagates inward as a
cooling wave (TD95). 
If the magnetic field is predominantly dipolar, then the radiative
flux across the field is concentrated near the surface of the star:  the
opacity of the E-mode scales as $B^{-2} \propto R^6$.  A cylindrical
bundle of field lines containing relativistic plasma therefore has a
radiative area (and luminosity) that decreases linearly with time,
as is observed in a number of short SGR bursts (e.g. Mazets et al.
1999a).
If higher multipoles dominate the near magnetic field, then the opacity
will be much more uniform over the surface of the fireball. 
Parametrizing
the radiative area in terms of the {\it remaining} fireball energy
as $A \propto E_{fireball}^{\alpha}$, the luminosity works out to
\begin{equation}
L_X(t) = L_X(0)\left(1-{t\over
t_{evap}}\right)^{\alpha/(1-\alpha)}.
\end{equation}
Here $t_{evap}$ is the time at which the cooling wave propagates to the
center of the fireball and the fireball evaporates.  This analytic law
provides an excellent fit to extended lightcurve of the August 27
event for $\alpha \sim 0.7$ (Feroci et al. 2000).

The fireball temperature inferred for the short SGR bursts is lower
if the confining volume is as large as in the giant
flares ($T \sim 100$ keV for an energy
$\sim 10^{41}$ erg).  However, a recent analysis of the August 29
burst from SGR 1900+14 (which appears to have been an aftershock
of the August 29 giant flare) points to an active region covering
only $\sim 0.1$ percent of the neutron star surface (Ibrahim et al. 2000).
This is consistent with a strong localization of the injected energy,
leading to the formation of a fireball with a similar temperature of
$\sim 1$ MeV.

%Alternative models for storing the energy of in the soft tail rapidly
%run into problems.   This energy ($1-3\times 10^{44}$ erg) exceeds what
%can be plausibly retained in an elastic deformation of the crust.  A
%strong core magnetic field ($B > 10^{15}$ G) could support a torsional
%Alfv\'en wave of this energy; but such a low frequency wave would excite the
%magnetosphere at a very large radius ($\sim 10^9$ cm), so large that
%it is difficult to see why the multipolar structure of the stellar
%field should be apparent in the lightcurve.

\subsection{QED Processes:  Radiative and Spectral Implications}

The transport of X-ray photons through a very strong (super-QED)
magnetic field is determined by two coupled processes:  Compton
scattering
and photon splitting $\gamma \rightarrow \gamma + \gamma$ (and
merging $\gamma + \gamma \rightarrow \gamma$) (TD95).  Even at very
large scattering depth, the dielectric properties of the medium are
dominated by vacuum polarization in the intense magnetic field. 
The normal modes of the electromagnetic field are then
linearly polarized (e.g. M\'esz\'aros 1992) with $\delta{\bf E} \perp
{\bf B}_0$ in one case (the extraordinary or E-mode) and
$\delta{\bf B} \perp {\bf B}_0$ in the other (the ordinary mode or
O-mode).

The E-mode splits because, in this situation, its energy and momentum
can be conserved by dividing it into two obliquely propagating O-mode
photons. (Or, at a lower rate, into a pair of E-mode and O-mode
photons.)
The O-mode is {\it not} able to split because its energy and momentum
cannot be so conserved.\footnote{These selection rules depend
essentially
on the inequality $n_O > n_E$ between the indices of refraction of the
two modes.  Note that both $n_O$ and $n_E$ depart only very
slightly from unity even in magnetic fields as strong as $\sim 10^{16}$
G.  The inequality is reversed, $n_E > n_O$, when plasma dominates the
dielectric properties of the medium; but in such a situation the
particle
density is enormous and the photons will in practice follow a Planckian
distribution.   The problem of calculating the emergent spectra of
SGR bursts focusses on much lower temperatures and densities where
departures from local thermodynamic equilibrium can occur.}
In a vacuum, neither mode is able to split for the simple reason that
the two daughter photons must remain colinear to conserve energy and
momentum, and there is no phase space for the process.  In marked
contrast with the strong $B^6$ scaling of the splitting rate in
sub-QED magnetic fields, the splitting rate approaches a $B$-independent
 value in fields much stronger than $B_{QED}$, 
$\Gamma_{sp}(\omega,B,\theta_{kB}) =
(\alpha_{em}^3/2160\pi^2)\,(m_ec^2/\hbar)\,(\hbar\omega/m_ec^2)^5\,
\sin^6\theta_{kB}$ (Adler 1971; Thompson \& Duncan 1992).  Here,
$\alpha_{em} \simeq 1/137$ and $\theta_{kB}$ is the angle between
the photon's wavevector and the background magnetic field.  This
implies immediately that a E-mode photon propagating a distance $R_{NS}
\sim
10$ km through a super-QED B-field will split if
$\hbar\omega > 38\,(R_{NS}/10~{\rm km})^{-1/5}$ keV (TD95; Baring 1995).

Compton scattering becomes strongly anisotropic in
a background magnetic field, with a strongly frequency-dependent
cross-section (M\'esz\'aros 1992).
In contrast with a dense plasma, both vacuum modes interact resonantly with an
electron (or positron) at the Landau frequencies. 
Near the surface of the star, the energy of the first Landau excitation
[$\simeq (2B/B_{QED})^{1/2}$ $m_ec^2$ when $B \gg B_{QED}$]
is much higher than the
temperature of the emerging X-radiation.  In this situation, there is
a strong suppression of the E-mode's scattering cross-section,
$\sigma_E = (\omega m_e c /eB_0)^2\,\sigma_T$, but not of the O-mode's
(e.g. Herold 1979).
This suppression greatly increases the radiative flux
close to the neutron star -- both from its surface (Paczy\'nski 1992,
Ulmer
1994) and across the confining magnetic field lines of a trapped
fireball
(TD95). 

However, even in the region where the E-mode is able to stream freely,
the O-mode with its large cross section
can still undergo many Compton scatterings and relax to a
Bose-Einstein distribution.   Given the strong frequency
dependence of the splitting rate,  there is clearly
a critical temperature above which the distributions
of the E- and O-modes both become thermal, which works out to
$T_{sp} = 11\,(R_{NS}/10~{\rm km})^{-1/5}$ keV (TD95).
This value agrees well with the best fit black body temperature
of the oscillatory tail in the August 27 giant flare (Feroci et al. 2000).
(The term `photon-splitting cascade'
appearing in some of the recent literature is a misnomer.  {\it If the Compton
depth is high enough to convert the O-mode back to the E-mode and allow
more than one generation of splitting, then X-rays are redistributed in
frequency mainly by the Compton recoil.})

The sharply peaked sub-pulses within the oscillatory tails of the
March 5 and August 27 events have a simple interpretation in the
magnetar model, and are consistent with the presence of a trapped
fireball (TD95).  This pattern requires a collimated, quasi-hydrodynamical
outflow of the X-radiation.  In an intense magnetic field, the rapid rise
of the E-mode opacity with radius provides a mechanism for
self-collimation:  if baryonic matter is suspended in the magnetosphere
by the hyper-Eddington radiative flux, then E-radiation will escape
by pushing this matter to the side.  In addition, a significant
fraction of the E-mode flux near
the E-mode photosphere is converted to the O-mode by scattering and by
photon splitting (Miller 1995; TD95).  The O-mode flows hydrodynamically
along the magnetic field even in the presence of a tiny amount of
matter,
$\dot M c^2/L_O \sim (GM_{NS}/R_{NS}c^2)^{-1}(L_O/L_{edd})^{-1}$.
Further collimation occurs if the photosphere is aligned
with extended (dipolar) magnetic field lines.

Scattering at the electron cyclotron resonance can probably be neglected
during outbursts from a magnetar:  the resonance sits at a large radius
$8\,R_{NS}$ $(B_{dipole}/10\,B_{QED})^{1/3}$
$(\hbar\omega/10\,{\rm keV})^{-1/3}$, where the outflowing photons are
sufficiently collimated to suppress the resonant scattering
depth below unity.  Scattering
at the ion cyclotron resonance has a significant optical depth $\tau_{ion}$ if
electrons
and ions dominate the {\it electron}-scattering opacity:
it is $\tau_{ion} \sim (\pi/4\alpha_{em})n_e\sigma_T R$$(B/B_{QED})^{-1}$
in a dipolar magnetic field.  An important
effect is to convert photons between the E- and O-modes and to
increase significantly the opacity of the E-mode at low frequencies.
(This may be relevant to the $< 7$ keV suppression of the radiative
flux found by Ulmer et al. 1993 in the bursts of SGR 1806-20.)

\subsection{Predictions of the Magnetar Model}

Here we mention three direct observational diagnostics of magnetars.

1) Afterglow radiation from the heated
surface that is exposed to a high temperature fireball.  The surface
absorbs $\sim 10^{-3}-10^{-2}$ of the fireball energy before it
dissipates.  This heat will be re-released on a timescale
comparable to or longer than the observed duration of the SGR outubrst.
The resulting luminosity increases monotonically with $B$, and is 
$\sim 10^{39}\times[{\rm exposed~area/(10~km)^2}]$
erg s$^{-1}$ for $T_{fireball} \sim 1$ MeV and $B \sim 10\,B_{QED}$ (TD95).
Direct evidence for afterglow at this level is present in a $\sim 4$
s burst from SGR 1900+14 that followed the August 27 flare by less
than 2 days (Ibrahim et al. 2000).  

2) Absorption or emission in the persistent emission at the ion 
cyclotron resonance
$\hbar\omega = 2.8\,(Z/A)(B/10~B_{QED})$ keV.  A direct measurement
of the surface magnetic field would be provided by the simultaneous
identification of a spin-flip transition:  the two frequencies
are very nearly degenerate for electrons, but differ by a factor
$2.8$ for protons.  This measurement is probably more feasible in
the AXPs, whose persistent emission has a much smaller non-thermal 
component than the SGRs.

3) Little or no reprocessing of X-rays into the IR/optical/UV bands by
an orbiting disk.  The persistent X-rays of the SGRs and AXPs do
not, in the magnetar model, result from accretion.  Only a modest
amount of material can be placed in orbit around the neutron
star through hyper-Eddington winds.  Detailed calculations of
reprocessing in the fossil disk model for the AXPs have been
performed by Perna, Hernquist, \& Narayan (2000)
(see also Chatterjee, Hernquist, \& Narayan 2000).  It is hard
to avoid reprocessing $\sim 10^{-2}$ of the central X-ray source
$L_X \sim 10^2\,L_\odot$ in this model.

\section{Variable Spindown of the SGRs and AXPs}

The four known SGRs are persistent X-ray sources, and in
two cases persistent periodicites  have
been detected:  $P = 7.47$ s for SGR 1806-20 (Kouveliotou et al.
1998);  and $P = 5.16$ s for SGR 1900+14 (Hurley et al. 1999c).
Together with the 8-s periodicity of
the March 5 event and the 6-12 s spin periods of the AXPs, these values
are clustered in a remarkably narrow range.  

\subsection{Spindown Ages}

In the magnetar model, the long spin periods of the SGRs were ascribed
to large torques driven by magnetic dipole radiation (DT92), possibly
amplified by a persistent flux of Alfv\'en waves and particles
(Thompson \& Blaes 1998).  A key motivation for this model came from the
early
association between the March 5 burster and the supernova remnant N49 in
the LMC (Cline 1982, and references therein):  the 8-s periodicity
corresponds to a magnetic dipole field
of $6\times 10^{14}$ G (polar) at an age of $\sim 10^4$ yr.
Of all the SGR and AXP sources, the spindown
of the AXP 1841-045 is most consistent with simple magnetic
dipole radiation (Gotthelf et al. 1999):  the spindown age $P/2\dot P
= 2000$ yr
agrees with the estimated age of the surrounding SNR Kes 73, and the
spindown
is very uniform.  The implied (polar) dipole field of $1.4\times
10^{15}$ G is a good candidate for the strongest yet measured in any
neutron star. 

However, the measured spindown ages of SGR 1806-20 and SGR 1900+14,
$P/2\dot P = 1400$ yr
(Kouveliotou et al. 1998, 1999;  Marsden, Rothschild, \& Lingenfelter
1999; Woods et al. 1999c) do not appear to obey a similar correspondence.
Indeed, the characteristic age of SGR 1900+14 is surprisingly short if
it has been spun down by a constant external torque, and if it
is physically associated with the nearby SNR G42.8+0.6:  the required
proper motion is $V_\perp \simeq 20,000\,(D/7~{\rm kpc})\,
(t/1,500~{\rm yr})^{-1}$ km s$^{-1}$.  (A spurious association leads to
an equally unsatisfactory situation: 
a very young neutron star bereft of a progenitor supernova.)
This inconsistency disappears if the spindown of SGR 1900+14 is {\it
temporarily accelerated} with respect to the long-term trend.

\subsection{DEPARTURES FROM UNIFORM SPINDOWN}

One of the defining characteristics of the AXPs is that they are
consistently spinning down.  Nonetheless, 
torque variations are evident in the sources
1E 2259+586, 1E 1048.1+5937, and 4U 0142+61 (Heyl \& Hernquist 1999;
Mereghetti et al. 2000).  Over the longest intervals measured, the 
spin evolution of the AXPs is unusually coherent by the standards
of accreting neutron stars.
Dramatically improved timing accuracy has been achieved recently
through phase-connected measurements (Kaspi et al. 1999), 
which show that the spindown of 
1E 2259+586 and RXSJ170849-4009 is, also, remarkably smooth over an interval
of $\sim 10^3$ days.  Thus, variations in the spindown may occur
only intermittently.  Recent phase-connected
measurements of the spindown of SGR 1806-20 (Woods et al. 2000)
point to significantly stronger torque variations 
than are present in the AXPs 1E 2259+586 and RXSJ170849-4009, or are
characteristic of radio pulsars.

Most intriguingly, the spindown of 1E 1048.1+5937 appears to 
have accelerated for a $\sim 5$ yr interval over the long term trend
(Paul et al. 2000), a phenomenon that was inferred for SGR 1900+14
only indirectly from its spindown age. 
The main question which next arises, is whether this accelerated
torque is comparable to that expected from an orthogonal vacuum rotator.
If so, then in the case of SGR 1900+14 the polar dipole field
is inferred to be $B_{dipole} \simeq 1\times 10^{15}$
G.  A temporary acceleration {\it up} to the standard dipole torque is
the expected consequence of a recent discharge of particles through
bursting activity, if the magnetic and rotational axes of the star are almost
aligned.  That is because in quiescence, the outer magnetosphere of the
neutron star would
become charge-starved at the measured 5.16-s spin period, with a corresponding
reduction in the long-term torque (T00).   Independent evidence for such
a torque reduction is present in the AXP 1E 2259+586,
whose spindown age of $\sim 2\times 10^5$ yr significantly {\it exceeds}
the estimated age of its host supernova remnant CTB 109.  

Also intriguingly, the spin period of SGR 1900+14 increased by
$\Delta P/P = +1\times 10^{-4}$ above the long term trend within an
80-day interval surrounding the August 27 giant outburst (Woods et al.
1999c).  A transient
flow of particles, photons, and Alfv\'en waves might provide the
additional torque -- by increasing the magnetic field strength at the
light cylinder and by carrying off angular momentum directly --
but the constraint on $B_{dipole}$ is severe (T00).  The net
effect of such a flow 
(Thompson \& Blaes 1998) is to increase the spindown luminosity
to the geometric mean of $L_{Alfven}$ and the standard magnetic dipole
luminosity, $I\Omega\dot\Omega  = \Lambda\, B_{NS} R_{NS}
(\Omega R_{NS}/c)^2\,(L_{Alfven}c)^{1/2}$.  Subsequent calculations have
found
the numerical coefficient to be $\Lambda = \sqrt{2}/3$ (Harding et al.
1999)
and $\Lambda = 2/3$ (T00).  Applying this formula to
the August 27 outburst, and normalizing the radiated
energy and duration to
the observed values ($\sim 10^{44}$ erg and $\sim 100$ s), one finds
$\Delta P/P = 1\times 10^{-5}\,(\Lambda/{2\over 3})\,
(\Delta E/10^{44}~{\rm erg})^{1/2}$
$(\Delta t/100~{\rm s})^{1/2}\,(B_{dipole}/10\,B_{QED})$.  This
falls below the measured value even for $B_{dipole} \sim 10\,B_{QED}$,
but a more extended particle flow or an undetected soft X-ray component
to the giant burst cannot be ruled out.

The long-term
spindown rate of SGR 1900+14 
appears not to have been perturbed by the August 27 event
(Woods et al. 1999c).  This observation has the important consequence
that the active region of the neutron star must carry a small fraction
of the external magnetic energy;  hence one deduces a lower bound to the
dipole field of $\sim 10\,B_{QED} = 4.4\times 10^{14}$ G (T00).

It should be kept in mind that the measured spindown luminosity
$I\Omega\dot\Omega$ of the AXPs and SGRs is typically two orders of 
magnitude smaller than the persistent X-ray luminosity.  
As the above formula makes clear, the spindown resulting from the
the release of a fixed energy increases with the duty cycle, because at
a lower
flux the Alfv\'en radius (and the lever arm) is increased.   
The inferred dipole fields of the two spinning-down SGRs
are reduced (by a factor of $\sim 4$) to $4\times 10^{14}$ G
if the each star is a persistent source of Alfv\'en waves and particles
with a luminosity comparable to $L_X \sim 10^{35}$ erg s$^{-1}$
(Thompson \& Blaes 1998; Kouveliotou et al. 1998, 1999; 
Harding, Contopoulos, \& Kazanas 1999; T00).  These values
lie only a factor $\sim 4$ above the polar dipole field
inferred for the new radio pulsar J1814-1744.  

The evidence for a physical association between SGR 1806-20
and a conspicuous radio nebula (Kulkarni
\& Frail 1993; Vasisht et al. 1994) which originally motivated persistent
particle winds from the SGRs 
has been called into question (Hurley et al. 1999d).
In addition, the relatively constant long-term spindown rate
of SGR 1900+14 (Woods et al. 1999c) indicates that transient surges in the
persistent seismic output, if present in SGR 1900+14, 
must be constant over a long timescale.  They do not appear to 
correlate directly with episodes of short outbursts.

Melatos (1999) has recently noted the intriguing possibility that the
spindown torque
coupled to the asymmetric inertia of the co-rotating magnetic field
could be
particularly effective at forcing precession in a magnetar.  Free
precession
(which in this model is modulated by the spindown torque) has a period
$\tau_{prec} = P/\varepsilon_B = 7\,(P/6~{\rm s})\,
(B_{core}/10^2 B_{QED})^{-2}$ day,
where $\varepsilon_B \simeq 1\times 10^{-5}(B_{core}/10^2 B_{QED})^2$
is the dimensionless quadrupole distortion of the star by the (toroidal
core) magnetic
field.  This period is already rather short compared with the observed
spindown variations, if the magnetic field is strong enough to power
the observed X-ray emission.  In addition, pinning of superfluid vortex
lines in the neutron star crust will force the precession period down
to $\tau_{prec} = P(I/I_{sf})$, where $I_{sf}$ is the fraction of the
moment of interia carried by the neutron superfluid (Shaham 1977).
Detection of a spinup glitch in 1RXS J170849.0-4000910 
provides direct evidence for sufficient pinning to suppress 
long-period precession (Kaspi, Lackey, \& Chakrabarty 2000).

\subsection{Superfluidity and Glitches}

Superfluid-driven glitches are a potential source of spindown
irregularities
in isolated magnetars.  SGRs 1900+14 and 1806-20 have
frequency derivatives about one-tenth that of the Vela pulsar, and
models
of magnetic dissipation suggest internal temperatures that are comparable
or higher (TD96; Heyl \& Kulkarni 1998).  A giant outburst like
the August 27 event must involve a large fracture of the crust
propagating
at $\sim 10^8$ cm s$^{-1}$, which almost certainly unpins the
$^1$S$_0$ neutron superfluid vortex lines
from the crustal lattice.   The maximum glitch that could result
can be very crudely estimated (TD96) by assuming a characteristic
maximum
angular velocity difference $\Delta\Omega_{max}$ between the superfluid
and lattice, and then scaling to the largest observed glitches
(e.g. $\Delta P/P \sim -3\times 10^{-6}$ in Vela).  This gives
$|\Delta\Omega/\Omega| \sim \Delta\Omega_{max}/\Omega \propto
\Omega^{-1}$ and
$|\Delta P/P| \simeq 3\times 10^{-4} (P/8~{\rm s})$.

In light of this, let us reconsider the observation of
a transient spindown in SGR 1900+14 ($\Delta P/P = 
+1\times 10^{-4}$) close to the
August 27 giant outburst (Woods et al. 1999c).   Could this be
a superfluid-driven glitch in spite of the `wrong' sign?   The
crust of a magnetar is deformed plastically by magnetic stresses
wherever $B >  (4\pi\mu)^{1/2} \sim 6\times 10^{15}$ G (TD96).  Such
a deformation taking place on a timescale short compared to $P/\dot P$
will force the pinned vortex lines into an inhomogeneous distribution
(with respect to cylindrical radius).  The net effect is to {\it slow}
the rotation of the superfluid with respect to the crust.  A sudden
unpinning event would then tend to {\it spin down} the rest of the star
(T00).

Heyl and Hernquist (1999) estimated the glitch activity in a few
variable AXPs, under the assumption that the spindown irregularities
are entirely due to glitches of the same sign as pulsar glitches. 
However,
recent phase-connected measurements of variable spindown in SGR 1806-20
do not support this hypothesis (Woods et al. 2000).
For that reason, it is important to consider alternative mechanisms
for spindown variations involving, e.g., acceleration of the torque
by particle outflows from the active neutron star.

\section{Persistent Emission of the SGRs and AXPs}

The persistent X-ray output of the Soft Gamma Repeaters lies within a
fairly narrow range of $1-10\times 10^{35}$ erg/s, and has a 
characteristically hard spectrum with a power-law component 
$dN/dE \propto E^{-2}$ (Murakami et al. 1994; Hurley et al. 1999c; 
Woods et al. 1999b).  
The output of the AXPs is slightly broader but much
softer spectrally, being well fit by a $\sim 0.5$ keV  blackbody plus
a (soft) non-thermal tail with photon index
$\sim 3-4$ (Mereghetti 2000, and references therein).  Direct evidence
that the persistent X-ray output of SGR 1900+14 is not powered by 
accretion comes from the detection of (enhanced) emission a day after
the giant August 27 flare (Murakami et al. 1999; Woods et al. 1999c).  
Enough radiative momentum
was deposited during that outburst to excavate any accretion disk out
to a very large radius, within which accretion would be established only
on a much longer timescale of months (T00).  

The spectral characteristics of the SGRs and AXPs suggest that
i) dissipation of magnetic energy in a neutron star can produce
varying
persistent X-ray spectra, with hardness correlating strongly with
bursting activity;
and ii) that more than one mechanism can generate persistent X-ray
emission at
a level of $\sim 10^{35}$ erg s$^{-1}$. Four such mechanisms have been
proposed, all of which can be expected to operate in an SGR and at least
two
of which are relevant to the AXPs.  We summarize them in turn:

{\bf 1.}  {\it Ambipolar diffusion of a magnetic field through the
neutron star core, combined with the increased transparency of the
stellar
envelope in a strong magnetic field} (TD96; Heyl \& Kulkarni 1998).
The degenerate charged electrons and protons are tied to the magnetic
field lines in the neutron star core.  They can be dragged across
the background neutron fluid, but only very slowly (Pethick 1992;
Goldreich \& Reisenegger 1992).
Heating of the core feeds back strongly
on the rate of ambipolar diffusion (TD96).  An intense magnetic field
drives an
imbalance between the chemical potentials of the electrons, protons and
neutrons,
$\Delta\mu = \mu_e + \mu_p - \mu_n \simeq B^2/8\pi n_e$, and this
imbalance
induces $\beta$-reactions which heat the core.  Above a critical flux
density,
the heat produced exceeds the heat remaining in the star from its
formation, and the
core sits at an equilibrium temperature where heating is balanced by
neutrino
cooling.  In practice, this balance is possible only as long as the
neutrino
emissivity is dominated by the modified-URCA reactions.  The very
strong temperature-dependence of these reactions translates into a
very strong $B$-dependence of the
diffusion rate:  $t_{amb} = 10^4\;(B_{core}/7\times 10^{15}~{\rm
G})^{-14}$ yr in a normal n-p-e plasma (TD96).
(This timescale depends, of course, on the {\it core} flux density.)
Assuming a magnetized iron envelope, the resulting heat flux through
the surface is $L_X(t) = 5\times 10^{34}\,(t/10^4~{\rm yr})^{-0.3}$
erg s$^{-1}$.  Thus, there is a critical flux density above which
magnetic dissipation is rapid, but below which the magnetic field
is essentially frozen.  {\it This critical flux density exceeds by
a factor $\sim 4-10$ the dipole fields that are inferred from SGR and
AXP spindown.}  The absence of significant X-ray emission from
the pulsar PSR J1814-1744 (Pivovaroff et al. 2000) 
(with a polar dipole field $\sim 10^{14}$ G) is consistent with the magnetar
hypothesis as long as the internal (e.g. toroidal) field in these
objects is below the critical value. 

Further time-dependent calculations of ambipolar diffusion
through normal n-p-e nuclear matter, including
much more detailed modelling of the envelope, are reported by
Heyl \& Kulkarni (1998).  Calculations of heat transport through
the strongly magnetized envelope of a neutron star are presented
in Hernquist (1985), Van Riper (1988), Potekhin \& Yakovlev (1996)
and Heyl \& Hernquist (1998).

{\bf 2.} {\it Hall fracturing in the crust} (TD96).
Protons are bound into a rigid Coulomb lattice of nuclei in the
neutron star crust.  In this situation, the propagation of
short-wavelength magnetic irregularities through the crust is driven
by the Hall electric field $\vec E = \vec J\times\vec B/n_e e c =
(\vec\nabla\times\vec B)\times\vec B/4\pi n_e e$ (Goldreich \& Reisenegger
1992).   The polarization of a such a Hall
wave rotates, which causes the crust to {\it yield or fracture} in a magnetic
field stronger than $\sim 10^{14}$ G.  A significant fraction of the
wave energy is dissipated in this manner -- in less than the age
$t_{NS}$ of the neutron star -- if the turbulence has a short wavelength
$\lambda < 0.1\,(\delta B/B)^{-1/2} (\theta_{max}/10^{-3})^{1/2}$
$(B^2/4\pi\mu)^{-1/4}\,(t_{NS}/10^4~{\rm yr})^{1/2}$ km (TD96).
(Recall that  $(4\pi\mu)^{1/2} = 6\times 10^{15}$ G in the deep
crust.) 
By contract, large-scale fractures which are capable of triggering
giant outbursts require more rapid transport of the magnetic field,
which can occur via ambipolar diffusion through the {\it core}.

Each Hall fracture releases only a small energy,
$\Delta E \sim 10^{36}\,(\theta_{max}/10^{-3})^{7/2}$ erg.  The
cumulative effect is to excite persistent seismic activity with a net
output
$L_{seismic} \sim 10^{35}\,(\delta B/B)^2\,
(B/10^{15}~{\rm G})^2$ $(t_{NS}/10^4~{\rm yr})^{-1}$
erg s$^{-1}$.
The excited seismic waves have a frequency
$\nu \sim c_\mu/\lambda \sim 10^4$ Hz, where $c_\mu \sim c/300$ is the
shear wave  speed in the deep crust.  These internal waves couple to
transverse (Alfv\'en) excitations of the magnetosphere at a radius
$R_\nu \sim c/3\nu \sim 100\,\lambda$.  Only a fraction of
the wave energy need be converted to particles to support the associated
electrical currents (Thompson \& Blaes 1998).
%Even in the absence of a broad spectrum
%of initial irregularities, large scale fractures can
%induce and modulate the emission of high frequency Hall waves
%by creating sharp gradients in the crustal magnetic field. 

{\bf 3.}  {\it Twisting of the external magnetic field lines by internal
motions of
the star, which drives persistent electrical currents through the
magnetosphere} (T00).
The persistent light curve of SGR 1900+14 underwent a dramatic change
following
the August 27 outburst (Murakami et al. 1999):  it brightened by a
factor $\sim 2.5$
and at the same time simplified dramatically into a single large pulse.
This change appeared within a day following the August 27 event,
indicating
that
the source of the excess emission involves particle flows {\it external}
to the
star (T00).  The coordinated rise and fall of the two X-ray pulses of 1E
2259+586
over a period of a few years detected by Ginga (Iwasawa, Koyama, \&
Halpern 1992) similarly indicates that some portion of its emission
is magnetospheric (TD96).

The rate of dissipation due to a twisting of a bundle of field lines
(of flux density $B$, radius $a$, twist angle $\theta$ and length
$L$)
can be estimated as follows (T00).  The associated charge flow is $\dot
N \sim
\theta B a^2 c/8 L$ into the magnetosphere from either end of the
twisted
field.
The surfaces of the SGRs and AXPs are hot enough to emit thermionically
for a wide range of surface compositions -- even in the presence of
$\sim 10^{15}$ G magnetic fields -- and so the space charge very nearly
cancels.
An electric field $\vec E\cdot \vec B =  -(Am_p/Ze)\vec g\cdot\vec B$
will
compensate the gravitational force on the ions; but the same field
pushes
the counterstreaming electrons to bulk relativistic motion.  The net
luminosity
in Comptonized X-ray photons is $L_{Comp} \sim 3\times 10^{35} \theta
(A/Z)
(B/10\,B_{QED})\,(L/R_{NS})^{-1}\,(a/0.5~R_{NS})^2$ erg s$^{-1}$.  This
agrees with the measured value if a few percent of the crust is involved
in the August 27 outburst.  (Independent evidence for an active fraction
this size comes from the unperturbed long-term spindown of SGR 1900+14,
and from the expectation of $\sim 10^2$ giant outbursts per SGR
in $\sim 10^4$ yr.)  This non-thermal energy source will decay in 10-100
years, and so provides a physical motivation for {\it non-thermal} persistent
X-ray spectra in {\it active} burst sources.   (Note that the measured
increase in the persistent $L_X$ of SGR 1900+14 came entirely in the
non-thermal component of the spectrum:  Woods et al. 1999a.)

{\bf 4.} Heyl and Hernquist (1997a,b) have
explored the interesting possibility that the emission of some AXPs is
predominantly due to {\it passive surface cooling, possibly enhanced by
a light H or He composition}.  This model is most promising for the AXP
1E 1841-045, but cannot directly 
accomodate the variable $L_X$ of 1E 2259+586 or
1E 1048.1+5937.  A challenge for this model comes from the very similar
spin periods and persistent X-ray luminosities of the active SGRs and
the
quiescent AXPs.  The magnetic dissipation occuring within an active SGR
lengthens its {\it lifetime} as an bright X-ray
source (TD96; Heyl \& Kulkarni 1998).

\subsection{Radio Emission from Magnetars}

The upper cutoff to the measured distribution of pulsar dipole fields
lies close to $B_{QED}$.  It is natural to ask whether this
apparent cutoff results from observational selection or, instead,
reflects a more fundamental physical limitation on pair cascades
in very strong magnetic fields.  As emphasized
by Camilo et al. (2000), the discovery of three new strong-field
pulsars allows for a substantial population of these objects.

The most salient point, I believe, is
that {\it particle flows induced by bursting activity} could
short out vacuum gaps in the magnetospheres of the SGRs.
The hyper-Eddington outbursts
can easily blow material off the stellar surface (TD95; Miller 1995;
Ibrahim et al. 2000).
A modest amount of material will remain centrifugally supported
outside the corotation radius (where the Keplerian orbital period 
equals the spin period of the star, 
$R_{co} = 6\times 10^8\,(P/6~{\rm s})^{2/3}$ cm)
and still be confined by the dipole magnetic field:
 $\Delta M \sim (B_{NS}^2/4\pi)\Omega^{4/3}R_{NS}^6$ $(GM_{NS})^{-5/3} =
3\times 10^{20}(B_{NS}/10\,B_{QED})^2\,
(P/6~{\rm s})^{-4/3}$ g.
After an X-ray outburst, this material
cools and settles into a rotationally supported disk, which spreads
outward adiabiatically as the centrifugal force density rises above
$B^2/4\pi$ at the co-rotation radius.  By spreading out to the
speed-of-light cylinder, this material can maintain a relativistic 
flux of particles at the Goldreich-Julian density 
$n_{GJ} = \Omega\cdot B/2\pi ec$ for as long as
$\sim 1\times 10^5\,(B_{NS}/10\,B_{QED})\,(P/6~{\rm s})^{2/3}$ yr.
Above this density, particles flowing downward toward the star 
from the light cylinder
would short out the `favorably curved' magnetic field lines 
that otherwise support a relativistic thermionic flow from the surface
of the neutron star (e.g. Scharlemann et al. 1978).  If this is
happening in the SGRs, radio emission may be easier to detect from the
quiescent AXPs.

Alternatively, it 
has been suggested that pair cascades (and thence coherent radio emission) 
are {\it directly} suppressed in isolated strong-field neutron stars 
through QED effects:  either by  conversion of gamma-rays
to bound positronium rather than to free pairs (Usov \& Melrose 1996);
or by photon splitting below the threshold for pair creation, 
$\gamma \rightarrow \gamma + \gamma$
(Baring \& Harding 1998).  Regarding the first mechanism, 
it should be emphasized that one polarization mode -- 
the E-mode -- converts to positronium with one particle in an
excited Landau state.  The 
energy released through the subsequent decay greatly exceeds the
binding energy of the pair, which suggests that an unbound pair results.
Notice also that the SGRs and AXPs emit
a large enough flux of soft X-rays ($L_X \sim 3\times 10^{34}-10^{36}$ 
erg s$^{-1}$) to photo-dissociate bound positronium (cf. Usov \&
Melrose 1996).   The other mechanism for suppressing radio emission
just discussed is based on a doubtful assumption that both 
polarization modes can split below the threshold for single-photon 
pair creation (see \S 4.5).  More generally,
the splitting rate drops off so rapidly with distance from 
the neutron star ($\propto B^6 \sim R^{-18}$ when $B < B_{QED}$)
that the region outside the `splitting photosphere' should sustain
a sufficient potential drop to induce pair cascades when
the spin period lies well below the conventional radio death line.  In other
words,  splitting will probably induce additional
curvature in the pulsar death line at $B > B_{QED}$, but not suppress
radio emission at much shorter spin periods.

\section{Origins of Neutron Star Magnetism}

The idea of magnetars was motivated by the realization that the
violent convective motions in a collapsing supernova core can strongly
amplify the entrained magnetic field (Thompson \& Duncan 1993, hereafter
TD93).
The intense flux of neutrinos drives convection both in the central part
of the core that is very thick to neutrino scattering and absorption
(Pons et al. 1999, and references therein) and in a thin mass shell
below
the bounce shock where neutrino heating overcomes cooling
(Janka \& Mueller 1996, and references therein).  Balancing
hydrodynamic and magnetic stresses, one deduces magnetic fields of
$\sim 10^{15}$ G and $\sim 10^{14}$ G respectively (TD93; Thompson
2000a).
The convection inside
the neutrinosphere has an overturn time $\tau_{con}$ of a few
milliseconds;
the overturn time in the outer `gain' region
is somewhat longer.  The inner region
will support a large-scale helical dynamo if the core is very rapidly
rotating, with $P_{rot} < \tau_{con}$ (DT92), but not otherwise.
It is also possible that rapid rotation by itself could
amplify a magnetic field (Leblanc \& Wilson 1970) through the magnetic
shearing instability (Balbus \& Hawley 1991) in the absence of
convection,
if the outermost parts of the collapsing core became centrifugally
supported.

A newborn neutron star experiences convection with
a dimensionless ratio of convective kinetic energy to gravitational
binding
energy ($\varepsilon_{con} \sim 10^{-4}$) that is some two orders of
magnitude larger than in any previous phase driven by nuclear burning
(TD93).
(This is the relevant figure of merit because the gravitational binding
energy and the magnetic energy are proportional under an expansion or
contraction.)  For this reason, neutron star magnetic fields are
probably not fossils from earlier stages of stellar evolution.
The intense flux of neutrinos emanating from the neutron core
induces rapid heating and $n-p$ transformations, thereby allowing
magnetic fields stronger than $\sim 10^{14}$ G to rise buoyantly
through a thick layer of convectively stable material in less than
the Kelvin time of $\sim 30$ s (Thompson \&  Murray 2000).  As a result,
the $10^{11}-10^{13}$ G magnetic moments of ordinary radio pulsars,
which do not appear to correlate with the axis of rotation,
have a plausible origin (TD93) in a stochastic dynamo operating
at slow rotation ($P_{rot} \gg \tau_{con}$).
Direct amplification of a magnetic field $\langle B^2\rangle^{1/2}$
within individual convective cells of size
$\ell \sim ({1\over 30}-{1\over 10})\,R_{NS}$ will generate a true
dipole
of magnitude
$B_{dipole} \sim \langle B^2\rangle^{1/2} (\ell^2/4\pi R_{NS}^2)^{1/2}
\sim 10^{13}$  G through an incoherent superposition.  A similar effect
can occur during fallback as convection develops below the
accretion shock (Thompson \& Murray 2000).

\subsection{Large Kicks}

There is evidence that some (but not all) SGRs have proper motions
approaching $\sim 1000$ km s$^{-1}$.  The quiescent X-ray source
associated with  SGR 0526-66 (the March 5 burster) is offset from the
center of N49, implying $V_\perp \sim 800\,(t/10^4~{\rm yr})^{-1}$ km
s$^{-1}$
perpendicular to the line-of-sight (DT92).  Similarly, the association
of SGR 1900+14 with G42.8+0.6, if real, implies $V_\perp \sim 3000\,
(t/10^4~{\rm yr})^{-1}$.  Additional indirect evidence that magnetars
tend to have received large kicks comes from the paucity of accreting,
strong-B neutron stars.   However, it should be emphasized that
the proper motions of SGR 1806-20 may be as small as $\sim 100$ km
s$^{-1}$,
and that the projected positions of the AXPs 1E 2259+586 and 1E 1841-045
are
close to the centers of their respective remnants. 

Since the SGRs already appear to have one unusal
property (very strong magnetic fields), one immediately asks if
a large kick could be produced by a mechanism that
does not operate, or operates inefficiently, in ordinary proto-pulsars.
Two mechanisms are particularly attractive if the star is initially
a rapid rotator (DT92; Khokhlov et al. 1999; Thompson 2000a):
anisotropy in the emission of the cooling
neutrinos caused by large scale magnetic spots, which suppress
convective
transport within the star; and asymmetric jets driven by late infall
of centrifugally supported material.  The first model is supported
by observations of rotating M-dwarfs, which have deep convective zones
(like proto-neutron stars) and develop large, long-lived polar magnetic
spots:  Vogt 1988).  One estimates
$M_{NS} V_{NS} \sim (E_\nu/c)\,(\tau_{spot}/\tau_{KH})^{1/2}\,
(\Delta\Omega_{spot}/4\pi)$,
where $\tau_{spot}$ is the coherence time of the spot(s) and $\tau_{KH}$
the Kelvin time of the star.  The corresponding magnetic dipole
field is  $B_{dipole} \sim 5\times 10^{14}\,
(V_{NS}/1000~{\rm km~s^{-1}})\,
(\tau_{spot}/\tau_{KH})^{-1/2}$ G.
Note that this refers to the dipole field in the {\it convective}
neutron
core, and represents an upper bound to the remnant field generated
by an internal dynamo.

An asymmetric jet provides a more efficient source of linear momentum
than
does radiation from an off-center magnetic dipole (Harrison \& Tademaru
1975),
for two reasons:  1) The jet is matter-loaded and the escape speed from
a proto-neutron star of radius $\sim 30$ km is only $\sim {1\over 4}$
the speed of light; and 2) a centrifigually supported disk carrying the
same amount of angular momentum $(GM_{core}R_{core})^{1/2}\Delta M$
as a hydrostatically supported neutron core can provide much more energy
to a directed outflow.  The respective energies are $\Delta E \sim
G M_{core}\Delta M/2R_{core} = 6 \times 10^{50}(\Delta
M/10^{-2}\,M_\odot)\,
(R_{core}/30~{\rm km})^{-1}$ erg and $\Delta E \sim
{5\over 4}G(\Delta M)^2/R_{core} = 10^{48}\,(\Delta
M/10^{-2}\,M_\odot)^2\,
(R_{core}/30~{\rm km})^{-1}$ erg for a $1.4\,M_\odot$ core.  The
corresponding kick velocity is $\sim 300\,f\,(\Delta
M/10^{-2}\,M_\odot)\,
(R_{core}/30~{\rm km})^{-1/2}$ km s$^{-1}$, where $f$ is the fractional
asymmetry in the momentum.  Only a very energetic jet
($\Delta E \sim 6\times 10^{51}\,(f/0.3)^{-1}$ erg) can generate a
kick of $1000$ km s$^{-1}$.

\acknowledgements
I thank Chryssa Kouveliotou, Shri Kulkarni, Peter Woods, and the late
Jan van Paradijs for many stimulating discussion about the peculiarities
of SGRs and AXPs;  Omer Blaes, Robert Duncan, and Norman Murray
for theoretical collaboration; and NASA (NAG 5-3100) and the 
Alfred P. Sloan foundation for financial support.

\end{document}